\documentclass[12pt]{article}
\pdfoutput=1
\usepackage{geometry,enumerate,amsmath,amssymb,amsfonts}
\usepackage{fullpage}
\usepackage{graphicx}
\numberwithin{equation}{section}
\newcommand{\be}{\begin{equation}}
\newcommand{\ee}{\end{equation}}
\newcommand{\bea}{\begin{eqnarray}}
\newcommand{\eea}{\end{eqnarray}}
\renewcommand{\hat}{\widehat}
\renewcommand{\tilde}{\widetilde}
\renewcommand{\epsilon}{\varepsilon}

\usepackage{cite}
\usepackage{amsmath,amssymb,amsfonts,graphicx}
\usepackage{epsfig}

\newcommand{\beq}{\begin{eqnarray}}
\newcommand{\eeq}{\end{eqnarray}}
\newcommand{\bec}{\begin{center}}
\newcommand{\eec}{\end{center}}
\newcommand{\M}{\mathbb{H}^3_\kappa}

\begin{document}
\title{
\vskip 10pt
\bf{Magnetic bags in hyperbolic space}
}
\author{\\[0pt] Stefano Bolognesi$^{\dagger}$, Derek Harland$^\star$ 
and Paul Sutcliffe$^\ddagger$\\[20pt]
{\em \normalsize $\dagger$ Department of Physics \lq\lq E. Fermi\rq\rq\,, University of Pisa,}\\
{\em \normalsize Largo Pontecorvo, 3, Ed. C, 56127 Pisa, Italy.}\\[10pt]
{\em \normalsize $\star$ School of Mathematics},\\
{\em \normalsize University of Leeds, Leeds LS2 9JT, U.K.}\\[10pt] 
{\em \normalsize $\ddagger$  Department of Mathematical Sciences,}\\ 
{\em \normalsize Durham University, Durham DH1 3LE, U.K.}\\[20pt] 
{\normalsize stefanobolo@gmail.com  \ d.g.harland@leeds.ac.uk \ p.m.sutcliffe@durham.ac.uk} 
}
\vskip 10pt
\date{April 2015}
\maketitle
\vskip 10pt
\begin{abstract}
  A magnetic bag is an abelian approximation to a large number of coincident
  $SU(2)$ BPS monopoles.
  In this paper we consider magnetic bags in hyperbolic space
  and derive their Nahm transform from the large charge limit of the
  discrete Nahm equation for hyperbolic monopoles.
   An advantage of studying magnetic bags in hyperbolic space, rather than
  Euclidean space, is that a range of exact
  charge $N$ hyperbolic monopoles can be constructed, for arbitrarily
  large values of $N$, and compared with the
  magnetic bag approximation.
  We show that a particular magnetic bag (the magnetic disc) provides a
  good description of the axially symmetric $N$-monopole.
  However, an abelian  magnetic bag is not a good approximation
  to a roughly spherical $N$-monopole that 
  has more than $N$ zeros of the Higgs field.
  We introduce an extension of the magnetic bag that does provide a good
  approximation to such monopoles and involves a spherical non-abelian
  interior for the bag, in addition to the conventional abelian exterior.
 
\end{abstract}
\newpage

\section{Introduction}\quad\
In three-dimensional Euclidean space
there is a $4N$-dimensional moduli space of $SU(2)$ charge $N$ BPS
magnetic monopole solutions of the Bogomolny equation.
If the $N$ monopoles are coincident, it has been proposed
that in the large $N$ limit
there is an abelian description, known
as a magnetic bag \cite{Bol}. This is a solution of the abelian
Bogomolny equation for a real scalar field, that approximates the
length of the Higgs field, and a $U(1)$ gauge field that models the
component of the non-abelian gauge field in the Higgs direction.
The bag is defined by a surface in $\mathbb{R}^3$ and the abelian fields
are taken to vanish in the interior of the bag.
Direct evidence for the magnetic bag description, in terms of a comparison
with the non-abelian fields of a monopole, is limited to 
low charge \cite{Bol,LW}, where a few axial and platonic monopole examples
are available \cite{book}. In particular, it has been observed that
the magnetic bag provides a reasonable
prediction for the size of these monopoles \cite{Man}. There is also
a monopole wall \cite{Ward}, with infinite magnetic charge, that 
resembles a local patch of the surface of a large magnetic
bag. 
Supporting evidence for the magnetic bag idea comes from the fact that
the Nahm transform \cite{Nahm} for monopoles becomes a transform
for magnetic bags in the large $N$ limit \cite{Har}.
Rigorous results relating to the size of a magnetic bag have recently been
obtained \cite{Tau}, and attempts have been made to numerically compute
non-abelian field configurations with large values of $N$,
with similar properties to a magnetic bag,
by gluing together cones of unit charge \cite{EG}.

The low charge platonic monopoles may be divided into two types,
by the structure of the zeros of the Higgs field \cite{Su4}.
The $N=4$ cubic monopole and the $N=7$ dodecahedral monopole have a single
zero of the Higgs field, with multiplicity $N$, at their centre.
This property is shared by the axially symmetric $N$-monopole,
 for all $N>1$.
Turning to the platonic solids with triangular faces, 
the tetrahedral, octahedral and icosahedral monopoles, with charges
$N=3,5,11$, have $N+1$ zeros of the Higgs field on the vertices of the
platonic solid and at their centre there is an additional zero with
multiplicity $-1$ (an anti-zero).
This led Lee and Weinberg \cite{LW} to propose that these low charge
monopoles are embryonic versions of large charge monopoles that can be
described by two extreme types of monopole bag,
which they named non-abelian and abelian bags respectively.
The first type models a monopole with a single zero of the Higgs field
(with multiplicity $N$) at the centre of the bag. The second type
describes a monopole that has most of the Higgs zeros
(in fact $N+1$ of them) distributed on the surface of the bag.
In this paper we shall have something to say about both types of monopole
bag, but the terms non-abelian and abelian are potentially
confusing given our later analysis. We therefore prefer to use the terms
cherry and strawberry flavour, to distinguish monopoles that have a 
large (in terms of multiplicity)
zero of the Higgs field at their centre from those
that have most of the Higgs zeros distributed on a surface. The nomenclature
is chosen because the distribution of the Higgs zeros mirrors the distribution
of the seeds in a cherry or a strawberry.

BPS monopoles in Euclidean space have a natural generalization to hyperbolic space, although a Nahm transform is known only if there is a discrete relationship between the curvature of hyperbolic space and the asymptotic length of the Higgs field. In this case hyperbolic monopoles
correspond to circle-invariant Yang-Mills instantons in $\mathbb{R}^4$ \cite{At} and are related to solutions of a discrete Nahm equation \cite{BA}.
In this paper we study magnetic bags in hyperbolic space and investigate their properties. 
We describe a transform that maps hyperbolic magnetic bags to solutions of a $u(\infty)$ Nahm equation and show how to derive this equation as the large $N$ limit of the discrete Nahm equation. 
If the asymptotic length of the Higgs field is suitably tuned then exact 
charge $N$ hyperbolic monopole solutions can be obtained in terms of free data specifying 
$N+1$ points on the sphere (together with a set of positive weights) \cite{MS}.
By taking large values of $N$ (we shall consider values of several hundred)
this provides large charge hyperbolic monopoles that can be used for
comparison with the magnetic bag approximation. This is a significant advantage
over the Euclidean situation.

Taking the points to be at the vertices of a regular $(N+1)$-gon, in an
equatorial circle on the sphere, yields the axially symmetric charge $N$
hyperbolic monopole. This monopole is cherry flavour, having a single
zero of the Higgs field of multiplicity $N$ at its centre.
In the large $N$ limit the associated magnetic bag is squashed into a
circular disc -- a magnetic disc. We compute an exact solution for the
magnetic disc and show that it provides a good approximation to the
axial $N$-monopole in the large $N$ limit.
If the $N+1$ points are sufficiently distributed over the sphere,
at the vertices of a deltahedron,
then the hyperbolic monopole is roughly spherical.
This monopole is strawberry flavour, with $N+1$ zeros of the Higgs field on the
vertices of the deltahedron and an anti-zero at the origin.
This is the
large $N$ generalization of the tetrahedral, octahedral and icosahedral
hyperbolic monopoles that arise from this construction with
$N=3,5,11$ \cite{MS}.
However, we find that the spherical abelian magnetic bag is not a good approximation to such hyperbolic $N$-monopoles, because the Higgs field
does not remain small inside the bag and also
has a significant spatial structure. 
We introduce an extension of the magnetic bag that applies when there are extra zeros of the Higgs field and show that this new bag does provide a good approximation to these large charge exact hyperbolic monopole solutions.
This sheds new light on the mysterious nature of monopole anti-zeros.

\section{Hyperbolic monopoles and magnetic bags}\quad\ 
In this section, we consider $SU(2)$ magnetic monopoles and bags
on three-dimensional hyperbolic space, $\M$, with constant curvature
$-\kappa^2$. The discussion is a straightforward generalization of the
Euclidean case $\mathbb{H}^3_0=\mathbb{R}^3$, and includes
this flat space limit.  We denote the metric on $\M$ by
\beq
ds^2(\M) = g_{ij}dx^i dx^j,
\label{genmetric}
\eeq
and its boundary by $\partial {\M}.$

The static energy of the $SU(2)$ Yang-Mills-Higgs theory is
\beq
E = \int_{\M}\ \left(- \frac{1}{8} {\rm Tr}\big( F_{ij}F^{ij} \big) -\frac{1}{4}  {\rm Tr}  \big( D_i\Phi D^i \Phi \big) \right)  \sqrt{g}\,d^3x,
\eeq
where $\Phi,A_i,$ are the $\mathfrak{su}(2)$-valued Higgs field and
the components of the gauge potential, with $F_{ij}=\partial_iA_j-\partial_jA_i+[A_i,A_j]$ and $D_i\Phi=\partial_i\Phi+[A_i,\Phi],$ for $i=1,2,3.$

The boundary condition on the Higgs field is that it has constant
positive magnitude $v$ at spatial infinity, that is
\be
|\Phi|^2=-\frac{1}{2}\mbox{Tr}(\Phi^2)=v^2 \qquad \mbox{on} \qquad \partial\M.
\ee
The monopole charge, $N\in\mathbb{Z}$, is given by the magnetic flux through
the boundary at infinity
\beq
N  = - \frac{1}{4 \pi v} \int_{\partial{\M}}  {\rm Tr}(F\Phi),
\eeq
where $F$ is the field stength two-form $F=\frac{1}{2}F_{ij}\,dx^i\wedge dx^j$.
To simplify the presentation, we shall restrict to the case $N>0.$
A standard Bogomolny argument yields the energy bound
\be
E\ge 2\pi v N,
\ee
which is attained by solutions of the first order Bogomolny equation
\beq 
F_{ij}  = \sqrt{g} \epsilon_{ijk} D^k\Phi.
\label{bog}
\eeq
As in flat space, there is a $4N$-dimensional moduli space of solutions
to (\ref{bog}), corresponding to arbitrary positions and $U(1)$ phases
for each of the $N$ individual monopoles.

For monopoles in hyperbolic space there are two length scales, namely,
the curvature scale of hyperbolic space $1/\kappa$, and 
the core size $1/v$ of a single monopole. The relevant
quantity is the ratio of these length scales, $v/\kappa$.
As first pointed out by Atiyah \cite{At}, if $2v/\kappa\in\mathbb{Z}$ then
a charge $N$ hyperbolic monopole is equivalent to a circle-invariant
self-dual Yang-Mills instanton in $\mathbb{R}^4$, with instanton number
$2Nv/\kappa.$ As a result, the study of hyperbolic monopoles simplifies
for discrete values of the asymptotic length of the Higgs field, relative
to the curvature of hyperbolic space.
As only the ratio is important, without
loss of generality we may choose to fix either $\kappa$ or $v$.
We shall choose to fix the former, by setting $\kappa=1$ from now on,
which means that the flat space limit is equivalent to the limit
$v\to\infty$ and the special tuned values for the length scale
are given by $2v\in\mathbb{Z}.$
For notational convenience we denote $\mathbb{H}_1^3$ by
$\mathbb{H}^3$. 

\renewcommand{\M}{\mathbb{H}^3}

A magnetic bag \cite{Bol} is an abelian approximation to a monopole solution
in the large $N$ limit, where all $N$ monopoles are coincident.
It involves a real scalar field $\phi$ and a $U(1)$ gauge field
$f_{ij} =\partial_i a_j - \partial_j a_i$, that are to be interpreted as
approximations to the length of the Higgs field and the projection of
the non-abelian gauge field onto the Higgs direction respectively
\beq
\label{identification}
\phi \approx |\Phi| \qquad  {\rm and} \qquad  f_{ij} \approx  - \frac{{\rm Tr}(F_{ij} \Phi)}{2 |\Phi|}.
\eeq
These abelian fields are required to satisfy the abelian Bogomolny equation
\beq
f_{ij} = \sqrt{g} \epsilon_{ijk} \partial^k \phi,
\label{abog}
\eeq
which implies that $\phi$ satisfies the Laplace-Beltrami equation
\beq
\partial_i (\sqrt{g} g^{ij} \partial_j \phi) = 0.
\label{lb}
\eeq
The magnetic bag is defined by specifying the surface of the bag
$\Sigma$, that divides
$\M$ into an interior and exterior part. In the interior of the bag
the abelian fields $\phi$ and $f_{ij}$ are taken to vanish.
The scalar field is required to vanish on the surface of the bag and
to have the correct asymptotic value at spatial infinity
\be
\phi=0 \quad \mbox{on}\quad \Sigma \qquad \mbox{and} \qquad \phi=v \quad
\mbox{on}\quad \partial\M.
\ee
Finally, the magnetic charge is identified with the abelian magnetic flux
through the surface of the bag
\be
N=\frac{1}{2\pi}\int_\Sigma f,
\label{aflux}
\ee
where $f=\frac{1}{2}f_{ij}\,dx^i\wedge dx^j$ is the abelian two-form field strength.

The idea is that the magnetic bag approximation improves with increasing $N$ and
becomes exact in the limit $N\to\infty,$ if accompanied by the limit
$v\to\infty,$ with $N/v$ non-zero and finite. This double scaling limit is
required to keep the size of the bag (and the hyperbolic $N$-monopole) finite
as $N\to\infty.$
Note that this limit does not correspond to the Euclidean limit, which
is $v\to\infty$ with $N/v\to 0.$

The freedom in choosing the surface $\Sigma$ reflects the fact that the
dimension of the $N$-monopole moduli space tends to infinity as $N\to\infty.$
The simplest example is the spherical bag, as follows.
We work with the ball model of hyperbolic space, given by the metric
\beq
ds^2(\M) = \frac{4}{(1-R^2)^2}\left( (dX^1)^2 + (dX^2)^2 + (dX^3)^2 \right),
\label{ballmetric}
\eeq
with radial coordinate $R = \sqrt{(X^1)^2 + (X^2)^2 +  (X^3)^2}<1$.
For a spherical bag, $\phi(R)$, the Laplace-Beltrami equation (\ref{lb})
reduces to
\be
\partial_R\bigg(\frac{R^2}{1-R^2}\partial_R\phi\bigg)=0.
\label{rlb}
\ee
Denoting the radius of the bag by $R_\star$, then $\phi(R)=0$ 
for $0\le R<R_\star$ and for $R\ge R_\star$ we require the solution
of (\ref{rlb}) that satisfies the boundary conditions $\phi(R_\star)=0$
and $\phi(1)=v.$ This solution is easily found to be
\be
\phi=\frac{v}{(1-R_\star)^2}\bigg(R_\star^2+1-\frac{R_\star}{R}(R^2+1)\bigg).
\label{rphi}
\ee
Substituting this solution into the abelian Bogomolny equation
(\ref{abog}) yields the abelian field strength, from which the magnetic charge
$N$ can be calculated using (\ref{aflux}). This provides the following
relation between the radius of the bag and the magnetic charge
\be
\frac{N}{v}=\frac{4R_\star}{(1-R_\star)^2}.
\label{bagradius}
\ee
This relation can be used to rewrite (\ref{rphi}) as
\be
\phi=v-\frac{N}{4R}(1-R)^2.
\label{rphi2}
\ee
This explicit example, and in particular the formula (\ref{bagradius}),
 illustrates the above discussion regarding
the double scaling limit, required to keep the size of the bag  finite
as $N\to\infty.$

It is helpful to rewrite the bag radius formula (\ref{bagradius})
in terms of the geodesic distance
from the origin $\rho=2\,\mbox{tanh}^{-1}R,$ to give
\be
\rho_\star=\frac{1}{2}\log\bigg(\frac{N}{v}+1\bigg).
\label{bagsize}
\ee
From this we see that if
the radius of the bag is much smaller than the curvature length scale,
$\rho_\star\ll 1$, then we recover the flat space result
$\rho_\star\approx N/(2v),$ that the bag radius grows linearly with
the magnetic charge. In contrast, for large bags $\rho_\star\gg 1$,
the radius has a
logarithmic growth with the magnetic charge.

For later use, we note that
in terms of the geodesic distance from the origin,
the expression (\ref{rphi2}) for the
scalar field of the spherical magnetic bag
is 
\be
  \phi=v(N+1-N\coth\rho).
 \label{outside}
  \ee

  A hyperbolic monopole is determined by the fields on $\partial\M$ \cite{BA},
  in contrast to Euclidean monopoles, where the fields on the
  sphere at infinity only fix the charge $N.$
  This distinction is also reflected in the magnetic bag description,
  because the surface of the bag $\Sigma$ is encoded
  in the abelian field strength on
  the boundary. To show this property, introduce spherical coordinates
  $R,\theta,\chi$ in the ball model of $\M$,
  \be
  X^1=R\sin\theta\cos\chi, \qquad\quad
  X^2=R\sin\theta\sin\chi, \qquad\quad
  X^3=R\cos\theta.
  \label{spherical}
  \ee
  As the scalar field $\phi$ of a magnetic bag is a harmonic function,
  it can be written as an expansion in terms
  of spherical harmonics $Y_{l,m}(\theta,\chi)$ as
  \be
  \phi=v-\frac{N}{4R}(1-R)^2+\sum_{l=1}^\infty\psi_l(R)\sum_{m=-l}^lc_{l,m}
  Y_{l,m}(\theta,\chi),
  \label{expansion}
  \ee
  where $\psi_l(R)$ is the solution of the ordinary differential equation
\be
  \partial_R\bigg(\frac{R^2}{1-R^2}\partial_R\psi_l\bigg)
  -\frac{l(l+1)}{1-R^2}\psi_l=0,
  \label{radialprofile}
  \ee
  satisfying the boundary condition
  \be
  \frac{\psi_l(R)}{(1-R)^2}\to 1 \quad \mbox{ as } \ R\to 1.
  \ee
  $\psi_l$ can be expressed in terms of an associated
  Legendre function of the first
  kind
  \be
  \psi_l(R)=\frac{(-1)^l}{(l+1)!}\sqrt{\frac{\pi(1-R^2)}{R}}
  P_\frac{1}{2}^{l+\frac{1}{2}}\bigg(\frac{1+R^2}{1-R^2}\bigg),
  \ee
  but we shall not need this explicit representation.

  It is clear from (\ref{expansion}) that all the expansion coefficients
  $c_{l,m}$ contribute to the computation of the surface of the bag
  $\Sigma$, given
  by $\phi=0.$ Substituting the expansion (\ref{expansion}) into the abelian
  Bogomolny equation (\ref{abog})
  and taking the limit $R\to 1$ yields the abelian field
  strength on the boundary sphere
  \be
  f=\bigg(\frac{N}{2}-2\sum_{l=1}^\infty\sum_{m=-l}^lc_{l,m}
  Y_{l,m}(\theta,\chi)\bigg)
  \sin\theta\, d\theta\wedge d\chi.
  \ee
  This shows that all the expansion coefficients $c_{l,m}$ contribute to the
  abelian field strength on $\partial\M$ and hence this contains the
  information required to reconstruct $\Sigma$.
  Note that all the coefficients $c_{l,m}$ vanish for a spherical bag,
  hence these coefficients provide a measure of the deviation of the
  bag from a spherical shape.

  \section{The hyperbolic  $\mathfrak{u}(\infty)$  Nahm equation}\quad\
  In Euclidean space there is a Nahm transform that relates magnetic bags
  to solutions of a $\mathfrak{u}(\infty)$ Nahm equation \cite{Har}.
  In this section, we describe
  a natural generalization of this transform to hyperbolic
  space.
  
  $\mathfrak{u}(\infty)$ is the Lie algebra of smooth real functions
  on $S^2$, with Lie bracket given by the Poisson bracket,
  and it may be regarded as the large $N$ limit of the Lie algebra
  $\mathfrak{u}(N)$ of hermitian $N\times N$ matrices
  \cite{Hop}.
  To be explicit, consider $S^2$ as the unit sphere in $\mathbb{R}^3$
  with cartesian coordinates ${\bf u}=(u^1,u^2,u^3)$.
  The standard area two-form on the sphere is given by
  $\omega=\frac{1}{2}\epsilon_{ijk}u^i\,du^j\wedge du^k$
  and the associated Poisson bracket is
  \be
  \{P,Q\}=\epsilon_{ijk}u^i\frac{\partial P}{\partial u^j}\frac{\partial Q}{\partial u^k}
  \ee
  for functions $P({\bf u}),Q({\bf u})$ on $S^2.$
  The algebra of functions on $S^2$ is generated by the
  cartesian coordinates, which clearly
    satisfy
     \be
     [u^i,u^j]=0 \qquad\mbox{ and }\qquad
     (u^1)^2+(u^2)^2+(u^3)^2=1,
     \label{functions}
     \ee
     together with the Poisson bracket relation
     \be\{u^i,u^j\}=\epsilon_{ijk}u^k.
     \label{pb}\ee
To reveal the connection to the large $N$ limit of $\mathfrak{u}(N)$, 
  let $J^1,J^2,J^3$ denote the generators of the $N$-dimensional irreducible
  representation of $\mathfrak{su}(2)$, satisfying
  $[J^i,J^j]=\epsilon_{ijk}J^k.$
  The algebra of hermitian $N\times N$ matrices is generated by
  $U^j=\frac{2i}{N}J^j$, satisfying
  \be
     [U^i,U^j]=\frac{2i}{N}\epsilon_{ijk}U^k \qquad\mbox{ and }\qquad
     (U^1)^2+(U^2)^2+(U^3)^2=1-\frac{1}{N^2}.
     \label{matrices}
     \ee
     The relations (\ref{matrices}) converge to the relations (\ref{functions})
     in the limit as $N\to\infty$, if we make the identification
     $U^j\to u^j$.
     Furthermore,
     in this limit the Poisson bracket relation (\ref{pb}) gives
       \be N[U^i,U^j]=2i\epsilon_{ijk}U^k\to
       2i\epsilon_{ijk}u^k=2i\{u^i,u^j\},
       \label{com2pb}
       \ee
       providing the prescription for replacing
       commutators by Poisson brackets.

       To define the hyperbolic $\mathfrak{u}(\infty)$  Nahm equation, let
       ${\bf x}=(x^1,x^2,x^3)$ be coordinates in $\M$ with metric
       (\ref{genmetric}) and consider the mapping
       \be
          {\bf x}:S^2\times [0,v)\mapsto \M,
            \ee
            defined by a solution of the equation
            \be \frac{dx^i}{ds}=\frac{\sqrt{g}}{N}g^{il}\epsilon_{jkl}\{x^j,x^k\},
            \label{hnahm}
            \ee
 where $s$ is the independent variable in the interval $[0,v).$         
   The boundary condition is that ${\bf x}$ is a coordinate on
   $\partial\M$ as $s\to v.$

   The Nahm transform for magnetic bags is simply an exchange of the
   independent and dependent variables in (\ref{hnahm}). The
   scalar field $\phi$ is identified with the variable $s$ and the
   abelian two-form $f$ is proportional to the area two-form $\omega$ on
   $S^2$,
   \be
   \phi=s, \qquad \qquad f=\frac{N}{2}\omega.
   \label{transform}
   \ee
   In particular, this identification means that ${\bf x}$ evaluated
   at $s=0$ is a 
   coordinate on $\Sigma$, the surface of the bag.
   
   In the Euclidean case, the proof that the
   $\mathfrak{u}(\infty)$ Nahm equation is equivalent to the
   abelian Bogomolny equation can be found in \cite{Har}.
 As    
  (\ref{hnahm}) is simply the covariant version of the
Euclidean equation, the proof follows from a simple covariant
version of the Euclidean proof. The main step is to multiply
(\ref{hnahm}) by $\omega\wedge ds$ and to use the property of the
Poisson bracket 
$\{x^j,x^k\}\,\omega\wedge ds=dx^j\wedge dx^k\wedge ds$
to see that $\frac{N}{2}\omega=*ds$,
where $*$ denotes the Hodge dual on $\M$.
This is the abelian
Bogomolny equation (\ref{abog}), given the identification (\ref{transform}).

As the Nahm transform linearizes the hyperbolic
$\mathfrak{u}(\infty)$ Nahm equation (\ref{hnahm}), then we expect this
to be an integrable system with an infinite number of conserved quantities.
In fact it is easy to show that
\be
\int\limits_{\{\phi=s\}}\Psi f,
\label{conserved}
\ee
is independent of $s$, for any harmonic function $\Psi$ on $\M$
with no singularities.
The proof is a simple application of Stokes' theorem, as follows,
\bea
&&\int\limits_{\{\phi=s_2\}}\Psi f \quad -\int\limits_{\{\phi=s_1\}}\Psi f \quad
=\int\limits_{\{s_1\le\phi\le s_2\}} d\Psi\wedge f \quad
=\int\limits_{\{s_1\le\phi\le s_2\}} d\Psi\wedge *d\phi \nonumber \\
&&=\int\limits_{\{s_1\le\phi\le s_2\}} d\phi\wedge *d\Psi \quad
=\int\limits_{\{\phi=s_2\}}\phi*d\Psi \quad
-\int\limits_{\{\phi=s_1\}}\phi*d\Psi \quad \\
&&=s_2\int\limits_{\{0\le\phi\le s_2\}} d*d\Psi \quad
-s_1\int\limits_{\{0\le\phi\le s_1\}} d*d\Psi \quad
=\ 0. \nonumber
\eea
In terms of spherical coordinates (\ref{spherical}), we may take
$\Psi=\tilde\psi_l(R)Y_{l,m}(\theta,\chi)$, where
\be
  \tilde\psi_l(R)=\frac{(l+1)!}{2}\sqrt{\frac{\pi(1-R^2)}{R}}
  P_\frac{1}{2}^{-l-\frac{1}{2}}\bigg(\frac{1+R^2}{1-R^2}\bigg)
  \ee
solves the radial equation (\ref{radialprofile}) and is 
normalized so that $\tilde\psi_l(1)=1.$
The conserved quantities (\ref{conserved}) are then proportional
to the constants $c_{l.m}$ that appear in the expansion (\ref{expansion})
of $\phi$.

To illustrate the Nahm transform for hyperbolic magnetic bags, we consider
the example of
the spherical bag, introduced in the previous section.
Using the ball model metric
(\ref{ballmetric}) the hyperbolic $\mathfrak{u}(\infty)$  Nahm equation
(\ref{hnahm}) becomes
\be
\frac{dX^i}{ds}=\frac{2}{N(1-R^2)}\epsilon_{ijk}\{X^j,X^k\}.
\label{ballhnahm}
\ee
In terms of the cartesian coordinates $u^i$ on $S^2$,
the spherically symmetric ansatz is given by
\be
X^i=u^i\,R(s).
\ee
Using the Poisson bracket relation (\ref{pb})
reduces (\ref{ballhnahm}) to the ordinary differential equation
\be
\frac{dR}{ds}=\frac{4R^2}{N(1-R^2)}.
  \ee
  The solution satisfying the required boundary condition, $R(v)=1$, is
  \be
  R(s)=1-\frac{2\sqrt{v-s}}{\sqrt{v-s}+\sqrt{N+v-s}}.
  \label{irphi2}
  \ee
  Setting $s=\phi$ in (\ref{irphi2}) indeed
  reproduces the spherical bag solution
  (\ref{rphi2}), in inverse function form.
  The bag radius is
  \be
  R_\star=R(0)=1-\frac{2}{1+\sqrt{1+N/v}},
  \ee
  which agrees with (\ref{bagradius}).
  
   \section{The large $N$ limit of the discrete Nahm equation}\quad\
   In Euclidean space, the $\mathfrak{u}(\infty)$ Nahm equation can be
   derived as a large $N$ limit of the 
   Nahm equation for $N\times N$ matrices \cite{Har}.
   This approach is not an option in hyperbolic space,
   as there is no known Nahm transform
   for generic values of $v$. However, for the tuned values $2v\in\mathbb{Z}$,
   there is a transform between hyperbolic monopoles and solutions of
   a discrete Nahm equation \cite{BA}. This lattice system is obtained
   by identifying hyperbolic monopoles with circle invariant instantons
   and imposing circle symmetry within the ADHM construction \cite{ADHM}.
   The lattice is indexed by the weight under the circle action and the
   construction yields a hyperbolic monopole within the upper half space
   model of $\M$, with metric
   \be
   ds^2(\M)=\frac{(dy^1)^2+(dy^2)^2+(dy^3)^2}{(y^3)^2},
   \label{uhs}
   \ee
   where $y^3>0$. The relation between the upper half space coordinates
   and the ball coordinates is
   \beq
y^3= \frac{1-R^2}{1+R^2-2X^3}, \qquad \qquad y^1+iy^2 = \frac{2(X^1 + i X^2)}{1+R^2-2X^3},
\label{ball2uhs}
\eeq
with the plane $y^3=0$ mapping to the boundary of hyperbolic space, $R=1$.

As a brief aside, note
this is the most convenient coordinate system in which to write down an
abelian magnetic wall, namely a solution of the abelian Bogomolny
equation that has translational
symmetry in a plane. Take $(y^1,y^2)$ to be the symmetry plane of the wall,
located at the position $y^3=y^3_\star$. The abelian fields vanish above the
wall ($y^3>y^3_\star$),  whereas below the wall ($0<y^3\le y^3_\star$) they
are given by
\be
\phi=v-v\bigg(\frac{y^3}{y^3_\star}\bigg)^2, \qquad\qquad\qquad
f=\frac{2v}{(y^3_\star)^2} \, dy^2\wedge dy^1.
\ee
We see that the magnetic flux is constant, hence the total magnetic flux
through the wall is infinite, as expected from the translational
symmetry. However, a finite piece of
this wall provides a good description of a local patch of the surface of a
large magnetic bag.

We now derive the hyperbolic $\mathfrak{u}(\infty)$ Nahm
equation from the large $N$ limit of the discrete Nahm equation.
       This discrete system is defined on a one-dimensional
       lattice consisting of $2v$ lattice points,
       $k=0,\ldots,2v-1$ with complex $N\times N$ matrices $B_{2j}$ and
       $W_{2j+1}$ defined on even and odd lattice sites respectively.
   For ease of presentation, we assume that $2v$ is an odd integer.
       The matrices are propagated along the lattice by applying the
       relations \cite{BA}
       \be
       B_{2j+2}=W^{-1}_{2j+1}B_{2j}W_{2j+1}
       \qquad\mbox{ and }\qquad
       W_{2j+1}W^\dagger_{2j+1}=W_{2j-1}W^\dagger_{2j-1}+[B_{2j}^\dagger,B_{2j}].
\label{dnahm}
       \ee
       Boundary conditions are introduced by extending the lattice to negative values
       and defining $B_{-2j}=B^t_{2j}$ and $W_{-(2j+1)}=W^t_{2j+1},$ plus adding
       an extra lattice site and demanding that $W_{2v}$ has rank one, so that
       $W_{2v}W^\dagger_{2v}=L^tL^\dagger$ for some $N$-component row vector $L.$

       The Nahm equation is obtained in the Euclidean flat space limit,
       $v\to\infty$, as follows \cite{BA}.
       Define the scaled lattice variable $\sigma=k/(2v)$ and write
       \be
       B_{2j}=-iT^1(\sigma)-T^2(\sigma)
       \qquad\mbox{ and }\qquad
       W_{2j+1}=v+T^3(\sigma+\frac{1}{2}v^{-1}).
       \label{bw2t}
       \ee
       There is a gauge symmetry of this system that allows $W_{2j+1}$ to
       be be chosen to be hermitian.
$\sigma$ becomes a continuous variable in
       the limit as $v\to\infty$ and the lattice system (\ref{dnahm}) becomes
       the Nahm equation \cite{Nahm}
       \be
       \frac{dT^i}{d\sigma}=-\frac{i}{2}\epsilon_{ijk}[T^j,T^k],
       \ee
       for the triplet of hermitian matrices $T^1,T^2,T^3$.

       The starting point to derive the large $N$ limit of the discrete Nahm
       equation is similar to the above.
       We introduce the same scaled lattice variable $\sigma$
       but we modify (\ref{bw2t}) by dropping the explicit $v$
       dependent term proportional to the identity matrix, to give
           \be
       B_{2j}=-iT^1(\sigma)-T^2(\sigma)
       \qquad\mbox{ and }\qquad
       W_{2j+1}=T^3(\sigma+\frac{1}{2}v^{-1}).
       \label{bw2t2}
       \ee   
       Substituting this form into the discrete Nahm equation (\ref{dnahm}),
       taking the large $v$ continuum limit
       and neglecting terms of order $v^{-1}$ yields
       \bea
       &&2\frac{dT^1}{d\sigma}T^3-\bigg[\frac{dT^1}{d\sigma},T^3\bigg]
       +iv\bigg[2T^2+\frac{1}{v}\frac{dT^2}{d\sigma},
         T^3+\frac{1}{2v}\frac{dT^3}{d\sigma}\bigg]=0\\
&& 2\frac{dT^2}{d\sigma}T^3-\bigg[\frac{dT^2}{d\sigma},T^3\bigg]
       -iv\bigg[2T^1+\frac{1}{v}\frac{dT^1}{d\sigma},
         T^3+\frac{1}{2v}\frac{dT^3}{d\sigma}\bigg]=0\\
       && 2\frac{dT^3}{d\sigma}T^3-\bigg[\frac{dT^3}{d\sigma},T^3\bigg]
       +2iv\bigg[T^1,T^2\bigg]=0.
       \eea
       Apply the large $N$ limit by replacing matrices by functions
       on the sphere
       $T^j(\sigma)\to y^j(\sigma,{\bf u})$,
       and using (\ref{com2pb}) to replace
       commutators by Poisson brackets,
       $[T^i,T^j]\to \frac{2i}{N}\{y^i,y^j\}$. This gives
   \bea
       &&\frac{dy^1}{d\sigma}y^3-\frac{i}{N}\bigg\{\frac{dy^1}{d\sigma},y^3\bigg\}
       -\frac{v}{N}\bigg\{2y^2+\frac{1}{v}\frac{dy^2}{d\sigma},
         y^3+\frac{1}{2v}\frac{dy^3}{d\sigma}\bigg\}=0\\
&& \frac{dy^2}{d\sigma}y^3-\frac{i}{N}\bigg\{\frac{dy^2}{d\sigma},y^3\bigg\}
       +\frac{v}{N}\bigg\{2y^1+\frac{1}{v}\frac{dy^1}{d\sigma},
         y^3+\frac{1}{2v}\frac{dy^3}{d\sigma}\bigg\}=0\\
        && \frac{dy^3}{d\sigma}y^3-\frac{i}{N}\bigg\{\frac{dy^3}{d\sigma},y^3\bigg\}
       -\frac{2v}{N}\bigg\{y^1,y^2\bigg\}=0.
       \eea     
       Finally, we take the limit $v\to\infty$ and $N\to\infty$ with ${v}/{N}$
       finite, to get
       \be
       \frac{dy^i}{d\sigma}y^3=\frac{v}{N}\epsilon_{ijk}\{y^j,y^k\}.
       \ee
       In the continuum limit 
       $\sigma\in[0,1),$ therefore to have the correct interval
         for the independent variable
         we introduce $s=v\sigma\in[0,v)$ to get the final form
   \be
       \frac{dy^i}{ds}=\frac{1}{Ny^3}\epsilon_{ijk}\{y^j,y^k\}.
       \ee
       This is the hyperbolic $\mathfrak{u}(\infty)$ Nahm equation
       (\ref{hnahm}) in upper half space coordinates with the metric (\ref{uhs}).
       The boundary condition on the discrete Nahm equation, that
       the rank of $W_{k}$ drops by a factor $1/N$ when $k=2v$,
       translates to the boundary
       condition that as $s\to v$ then $y^3\to 0$,
       which is indeed the boundary of hyperbolic space, in upper half space
       coordinates.

       \section{Exact hyperbolic monopoles with large charge}\quad\
By restricting to the simplest tuned value, $v=\frac{1}{2}$,
explicit exact charge $N$ hyperbolic monopole solutions can be obtained
from free data specifying 
$N+1$ points on the sphere (together with a positive weight for each point) \cite{MS}.
At the heart of this construction is the identification of a hyperbolic
$N$-monopole with a circle-invariant $N$-instanton in $\mathbb{R}^4$
obtained using the JNR ansatz \cite{JNR} for instantons, with JNR poles
restricted to the fixed point set of the circle action.
An alternative view of the same solution is via the discrete Nahm equation
discussed in the previous section, where the restriction $v=\frac{1}{2}$
reduces the lattice to a single point. All that remains of the discrete Nahm
equation is then a boundary condition for the complex 
$N\times N$ symmetric matrix $B_0$ and the
complex row vector $L$ that gives $W_1$.
The solution associated with the free data is essentially obtained by taking
$B_0$ to be diagonal, with the remaining
data providing the components of $L$ in a simple way that
automatically satisfies the boundary condition \cite{BCS}.  

An explicit formula for the Higgs field is most naturally written using the
upper half space coordinates (\ref{uhs}), no matter whether the JNR or discrete
Nahm route is taken to obtain the solution. To present this formula, 
let  
$\{\gamma_j\in\mathbb{CP}^1,j=0,\ldots,N\}$ be a set of $N+1$ distinct points
on the Riemann sphere and use these points to define the following real
function
       \be
       \Xi=\sum_{j=0}^N \frac{1+|\gamma_j^2|}{|y^1+iy^2-\gamma_j|^2+(y^3)^2}. \label{JNR}\ee
       The square of the length of the Higgs field is then given by \cite{MS,BCS}
       \be
|\Phi|^2=\bigg(\frac{y^3}{2\Xi}\bigg)^2\bigg(
\bigg(\frac{\partial\Xi}{\partial y^1}\bigg)^2+
\bigg(\frac{\partial\Xi}{\partial y^2}\bigg)^2
+\bigg(\frac{\Xi}{y^3}+\frac{\partial\Xi}{\partial y^3}\bigg)^2\bigg).
\label{higgs}
\ee
Although this formula for the Higgs field is most readily obtained
in upper half space coordinates,
the symmetry of the solution is most apparent by converting to
the ball model using the relations (\ref{ball2uhs}) between the
two coordinate systems. This reveals that the points $\gamma_j$ on the
Riemann sphere should
be regarded as points on the sphere $R=1$, that is the boundary of
$\M$ in the ball model. Furthermore, the monopole inherits the
symmetry of this set of points on the sphere,
due to the choice of weights in (\ref{JNR}).
Replacing the weights $1+|\gamma_j|^2$ in (\ref{JNR}) with
arbitrary real and positive weights also yields a hyperbolic
monopole solution, but generally this will not share the symmetry of the
set of points on the sphere.

The axially symmetric hyperbolic $N$-monopole (positioned at the origin,
with $X^3$ the axis of symmetry) is obtained in this
formalism by the choice $\gamma_j=e^{2\pi ij/(N+1)}.$
Naively, it might be expected that placing $N+1$ points
on the vertices of a regular $(N+1)$-gon in an equatorial circle would
produce a monopole with a discrete cyclic symmetry, but the fact that all
the points lie on a circle enhances the cyclic symmetry to an axial symmetry.
For later reference, in the plane $X^3=0$ the length of the
Higgs field has the simple
expression \cite{Co}
\be
|\Phi|=\frac{(N+1)R^N(1-R^2)}{2(1-R^{2N+2})}.
\label{axial}
\ee
From this formula we see that the axial $N$-monopole indeed has a
zero of the Higgs field at the origin, with multiplicity $N$.
This means that the axial $N$-monopole is cherry flavour. 

\begin{figure}
\begin{center}
\includegraphics[width=7cm]{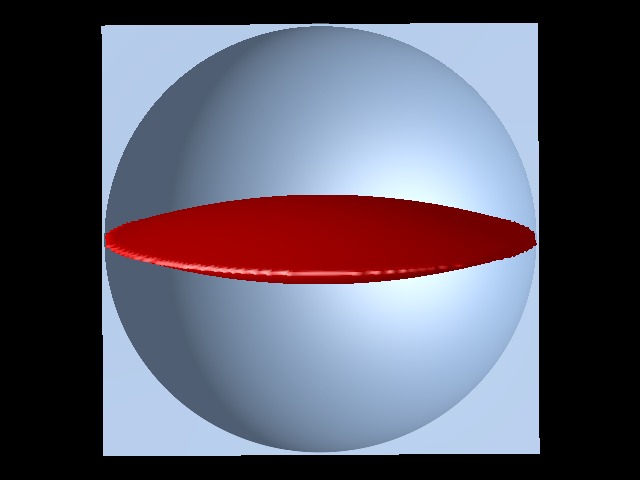}
  \includegraphics[width=7cm]{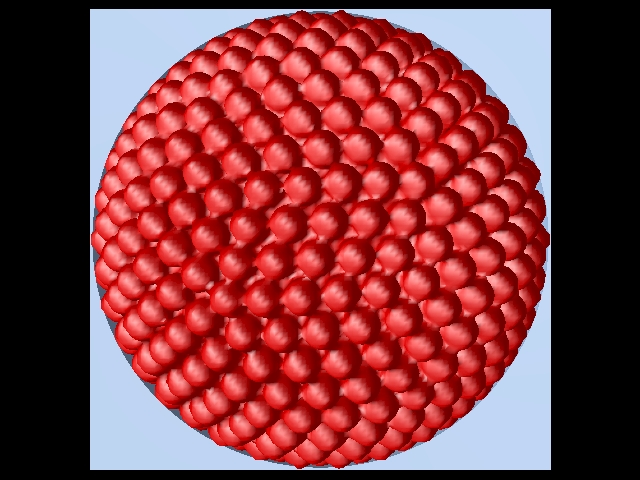}
  \caption{Energy density isosurfaces, in the ball model of $\M$, for
    hyperbolic monopoles with $N=371$. The hyperbolic monopole
    in the left image has axial symmetry and is cherry flavour,
    whereas the one in the right image has icosahedral symmetry
    and is strawberry flavour.
    The blue sphere represents the boundary of hyperbolic space.}
\label{fig-371} 
\end{center}
\end{figure}

The energy density of a monopole solution can be obtained directly from
the length of the Higgs field by acting with the Laplace-Beltrami operator on
$|\Phi|^2.$ In the left image in Figure~\ref{fig-371} we display an
energy density isosurface, using the ball model of $\M$,
for the axial monopole with $N=371$ (the
reason for this particular choice of $N$ will be revealed shortly).
The blue sphere in this image represents the boundary of hyperbolic space, $R=1$. 
We see that, for a large value of $N$, the energy density isosurface
of the axial $N$-monopole takes the form of a thin disc. In the next section
we study the magnetic bag approximation to this type of solution, namely
the magnetic disc, and show that it provides a good description.

Applying the above construction with $N=3,5,11$ and placing the $N+1$ points
on the sphere at
the vertices of a tetrahedron, octahedron and icosahedron, respectively,
yields a tetrahedral 3-monopole, an octahedral 5-monopole and an icosahedral
11-monopole \cite{MS}. All these monopoles are strawberry flavor, with an
anti-zero at the origin and $N+1$ zeros of the Higgs field on the vertices
of a platonic solid.
We can continue this family to large $N$, by placing the $N+1$
points on the vertices of a suitable deltahedron, so that the points
are in some sense evenly distributed.
Although there are no spherically symmetric monopoles with $N>1$,
within the moduli space of monopoles obtained from the free data of points
on a sphere,
this family generates an $N$-monopole that is the best 
candidate to have a spherical abelian bag description.

To generate $N+1$ evenly distributed points on the sphere we turn to
the following
well-known physical problem. Given a positive integer $M$,
the Thomson problem is to find the positions
of $M$ unit charge point particles on the sphere that attain the
global minimum of their total
electrostatic Coulomb energy (for a review see \cite{EH}).
For $M=4,6,12$ the solution of the Thomson problem is to place the
point particles at the vertices of a tetrahedron, octahedron and
icosahedron respectively. By taking our points on the sphere
to be the positions of the particles that solve the Thomson problem
we can generate a family of hyperbolic monopoles with charge $N=M-1$
that includes and extends our platonic strawberry flavour examples.

Computing solutions of the Thomson problem for large $M$ is a
difficult computational task, due to the large number of local
minima that exist. However, this is a well-studied optimization problem,
that is often used to benchmark new algorithms, so there is a wealth of
data available. In particular, magic numbers have been found at which
icosahedrally symmetric local energy minima have been obtained that are
believed to be the global minima. Icosahedral symmetry is the best approximation
to spherical symmetry that can be obtained with a finite number of points,
hence this is the closest that we can come to a spherical configuration.
As an example, it is believed that $M=372$ is a magic number with
icosahedral symmetry \cite{WU}. Taking this configuration of points
yields the icosahedrally symmetric
hyperbolic monopole with charge $371$ displayed
in the right image in Figure~\ref{fig-371}. This explains our earlier
non-obvious choice of $N=371$ for the axial monopole, as we want to
display the energy density isosurfaces of the two different kinds of
monopole with the same charge, to aid the comparison.

An examination of the Higgs field of the
icosahedrally symmetric 371-monopole displayed in the right image in
Figure~\ref{fig-371} confirms that this is indeed strawberry flavour,
with an anti-zero at the origin and 372 zeros on a shell. In section
\ref{strawberry} we shall discuss the Higgs field of this monopole
in detail and explain why an abelian magnetic bag is not a good description.
We then introduce a new magnetic bag with a non-abelian interior that
does provide a good approximation to strawberry flavour monopoles.

Finally in this section, we stress that we expect there to be
a family of charge $N$ cherry flavour hyperbolic monopoles that approach the
spherical abelian magnetic bag in the large $N$ limit. This is a family
that extends the cubic 4-monopole and the dodecahedral 7-monopole,
obtained by imposing constraints on the ADHM construction that
ensures a circle symmetry of the instanton \cite{MS}.
These examples are
not within the scheme of specifying free data as points on a sphere
and hence we are currently unable to extend the family to large values
of $N$ because of the technical difficulty in imposing the required
constraints.

\section{The magnetic disc}\quad\
A magnetic disc is the degenerate limit in which the surface $\Sigma$
of the magnetic bag becomes a disc.
Therefore, to obtain a magnetic disc we 
require a harmonic function that vanishes on a disc. It is possible
to obtain the required solution explicitly by introducing an appropriate
coordinate system, in terms of Jacobi elliptic functions, with the
property that the Laplace-Beltrami equation has solutions that can be
obtained via a separation of variables \cite{Ka}.

Consider the disc, ${\cal D}_S$,
of geodesic radius $S$, given in ball coordinates
by $X^3=0$ and
$\sqrt{(X^1)^2+(X^2)^2}\le{\rm tanh}(S/2)$.
Let ${\rm sn}$ denote the Jacobi elliptic function with elliptic
modulus
${\rm tanh}S$ and $\widetilde{\rm sn}$ the Jacobi elliptic
function with elliptic modulus ${\rm sech}S$. We extend the same notation to the
other Jacobi elliptic functions and to the complete elliptic
integral of the first kind, so that $K$ denotes this elliptic integral
with elliptic modulus ${\rm tanh}S$ and $\widetilde K$ is the
 complete elliptic integral of the first kind with 
 elliptic modulus ${\rm sech}S$.

 We introduce the coordinates $r,\Theta,\chi$ on $\M$, where
 $0\le r < \widetilde K$ and the angular coordinates have the
 ranges
 $-K<\Theta<K$ and $0\le\chi\le 2\pi.$
 The relation to the ball coordinates is given by
\bea
X^1&=&\frac{{\rm sinh}S\, \widetilde{\rm nc}(r)\,
  {\rm cn}(\Theta)\,\cos\chi
}{1+{\rm cosh}S\, \widetilde{\rm dc}(r)\,
  {\rm dn}(\Theta)},\\
X^2&=&\frac{{\rm sinh}S\, \widetilde{\rm nc}(r)\,
  {\rm cn}(\Theta)\,\sin\chi
}{1+{\rm cosh}S\, \widetilde{\rm dc}(r)\,
  {\rm dn}(\Theta)},\\
X^3&=&\frac{{\rm tanh}S\, \widetilde{\rm sc}(r)\,
  {\rm sn}(\Theta)
}{1+{\rm cosh}S\, \widetilde{\rm dc}(r)\,
  {\rm dn}(\Theta)},
\eea
and yields the metric
\be
ds^2(\M)=\big( \widetilde{\rm dc}^2(r)-{\rm dn}^2(\Theta)\big)(dr^2+d\Theta^2)
+{\rm sinh}^2S\, \widetilde{\rm nc}^2(r)\,{\rm cn}^2(\Theta)
d\chi^2.
\ee

\begin{figure}
\begin{center}
\includegraphics[width=6cm]{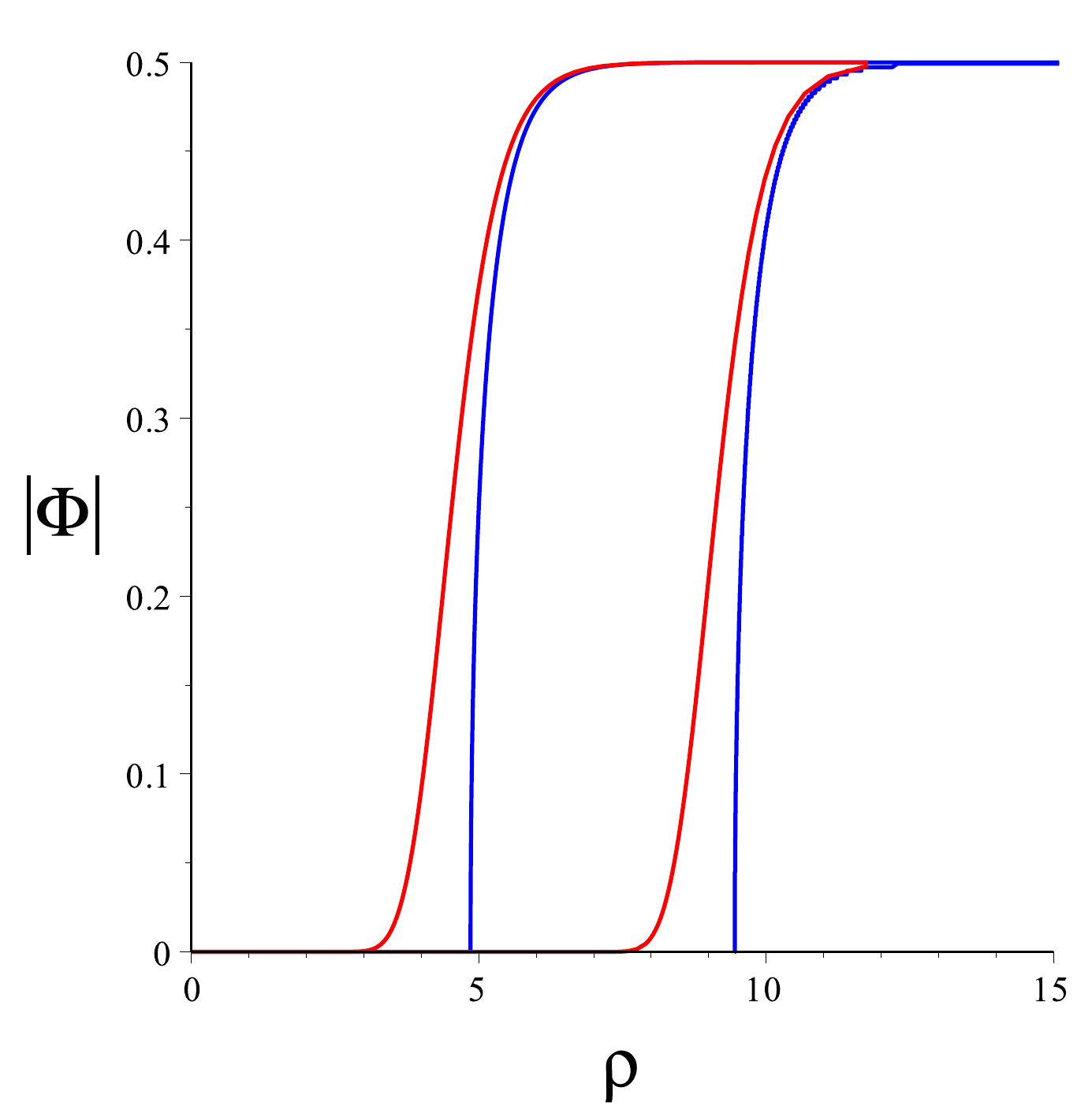}
\caption{The red curves display the length of the Higgs
  field $|\Phi|$ as a function of geodesic
  distance from the origin $\rho$, along an axis that is perpendicular
  to the symmetry axis of the axial hyperbolic $N$-monopole with
  $N=100$ and $N=10000$. The blue curves show the corresponding magnetic
disc approximation.}
\label{fig-2disks} 
\end{center}
\end{figure}

The first reason for using this coordinate system is that the disc ${\cal D}_S$
is simply given by $r=0$. The second reason is that this allows
a separable solution of the Laplace-Beltrami equation (\ref{lb})
with $\phi$ a function of $r$ only.
The ansatz $\phi(r)$ reduces (\ref{lb}) to the ordinary
differential equation
\be \frac{d}{dr}\bigg( \widetilde{\rm nc}(r)\frac{d\phi}{dr}\bigg)=0.\ee
We require the solution that vanishes on the disc ${\cal D}_S$, hence $\phi(0)=0$,
and has the correct asymptotic value,
$\phi(r)\to v$ as $r\to  \widetilde K.$
As we wish to compare the magnetic disc with the axial exact solution in
the previous section we take $v=\frac{1}{2}$, so the required 
solution is
\be
\phi(r)=\frac{\cos^{-1}( \widetilde{\rm dn}(r))}{2\cos^{-1}({\rm tanh}S)}
\label{phidisc}
\ee
The relation between the magnetic charge and the geodesic
radius of the disc is given by
\be
N=\frac{1}{2\pi}\int_{\Sigma_r} *d\phi=\frac{\sin^{-1}({\rm tanh}S)}{\cos^{-1}({\rm tanh}S)},
\ee
where $\Sigma_r$ is any surface of constant $r$.
Inverting this formula provides the geodesic radius of the disc
\be
S=\tanh^{-1}\bigg(\sin\bigg(\frac{\pi N}{2(N+1)}\bigg)\bigg)
=\log N+\log\bigg(\frac{4}{\pi}\bigg)+{\cal O}\bigg(\frac{1}{N}\bigg).
\label{discradius}
\ee
  Along the positive $X^1$ axis, in the exterior of the disc,
  the relation between
  $X^1=R$ and the coordinate $r$ is
  \be
  R=\frac{{\rm sinh}S\, \widetilde{\rm nc}(r)
  }{1+{\rm cosh}S\, \widetilde{\rm dc}(r)}.
  \ee
Using this formula, in Figure~\ref{fig-2disks} we plot the solution (\ref{phidisc}) 
as a function of the geodesic distance from the origin
$\rho=2\tanh^{-1}R$, for the charges
$N=100$ and $N=10000$ (blue curves).
For comparison, the red curves in Figure~\ref{fig-2disks}
display the corresponding exact solution
(\ref{axial}) along the same axis, again as a function of $\rho$.
We see that the magnetic disc provides a reasonable approximation
to the exact axial monopole and that the error appears to have
very little dependence on $N$ for these large values. As we now explain, this
is exactly the result expected of a magnetic bag approximation.

  To compare the disc radius (\ref{discradius}) with the exact
  axial monopole solution, we use (\ref{axial})
  to define the value $\hat R$ at which the Higgs
  field attains half the asymptotic value,
  $|\Phi|=\frac{1}{4}$.
This provides a sufficient definition of the size of the axial monopole.
  The geodesic radius of the axial monopole is
then given by
\be
\hat S=2\tanh^{-1}(\hat R)=\log N+\log \bigg(\frac{2}{\sqrt{3}}\bigg)
+{\cal O}\bigg(\frac{1}{N}\bigg).
\label{axialradius}
\ee
Comparing (\ref{discradius}) and (\ref{axialradius}) shows that the two
agree up to terms that are ${\cal O}(1)$. Recall that the magnetic bag
is expected to become exact in the limit $N\to\infty$ and $v\to\infty$
with $N/v$ finite. The $v\to\infty$ limit is required to keep the size
of the magnetic bag finite. However, our exact solutions are only available
for $v=\frac{1}{2}$, so we are unable to take the $v\to\infty$ limit to
keep the size finite as $N\to\infty.$ An alternative is to measure
geodesic distance in units of $\log N$, so that, by
(\ref{discradius}), the magnetic disc
has geodesic radius one in these units as $N\to\infty$.
In these units, terms that are ${\cal O}(1)$ tend to zero
as  $N\to\infty$, and hence the exact axial monopole converges to
the magnetic disc.

Note that (\ref{bagsize}) shows that
in the large $N$ limit, with $v=\frac{1}{2}$, the
leading order term for the geodesic radius of the spherical bag is
$\log\sqrt{N}$, in comparison to the geodesic radius of the
magnetic disc, $\log N$. Thus the spherical bag is a
substantially more compact object
than the magnetic disc.

  \section{A magnetic bag for strawberry flavour monopoles}\label{strawberry}\quad\
  In the previous section we considered a particular type of cherry
  flavour monopole, the axial monopole, and demonstrated that the
  abelian magnetic bag indeed provides a good description in the large charge
  limit.
  In this section we turn our attention to strawberry flavour
  monopoles and find that the abelian magnetic bag is no longer a
  good approximation.
  
\begin{figure}
\begin{center}
\includegraphics[width=11cm]{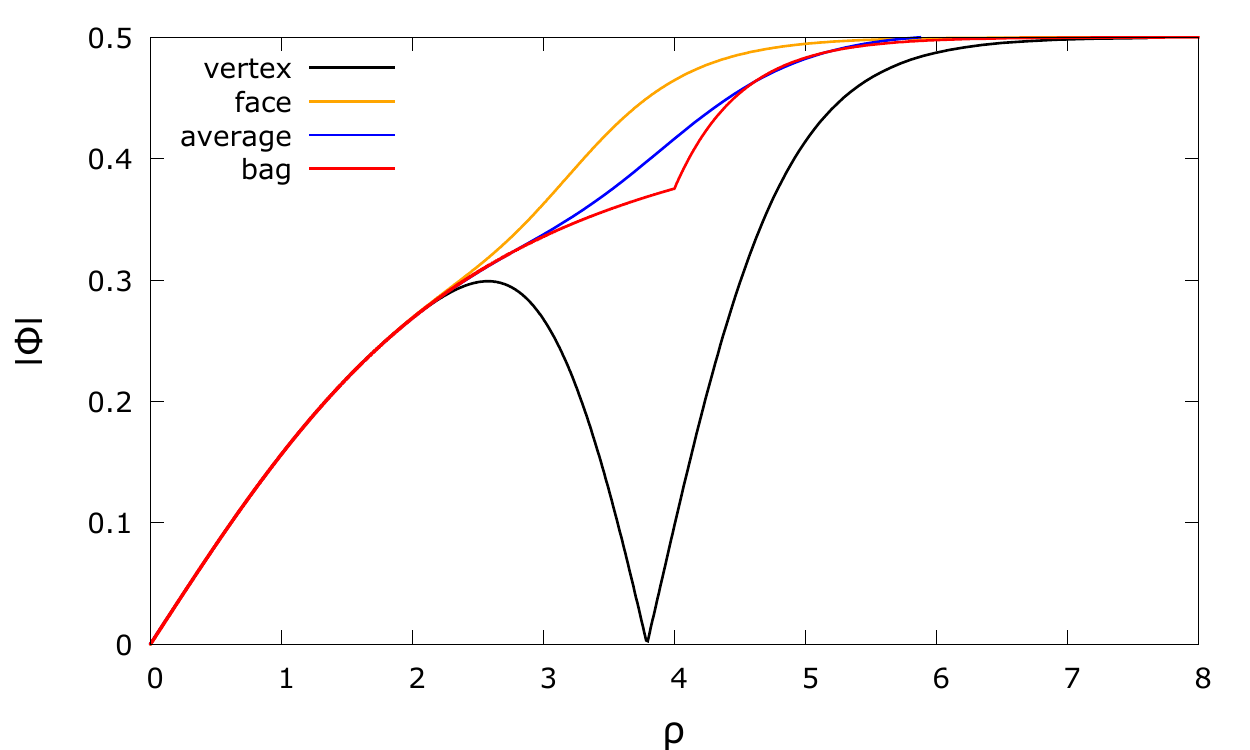}
\caption{The length of the Higgs field, $|\Phi|$, for an
  icosahedrally symmetric strawberry flavour monopole with
  charge $N=371.$ The plot shows $|\Phi|$ as a
 function of geodesic distance from the
  origin $\rho$ along a radial half-line that passes through a
  vertex (black curve) and a face centre (yellow curve) of the
  associated polyhedron. The blue curve is the
  spherical average of
  $|\Phi|$, obtained by integrating over the angular coordinates. 
The red curve is the new magnetic bag approximation.
}
\label{fig-antibag} 
\end{center}
\end{figure}

  A typical example of a large charge strawberry flavour hyperbolic
  monopole is the 
  icosahedrally symmetric charge 371 monopole displayed in the right
  image in Figure~\ref{fig-371}.
  This has an anti-zero at the origin and 372 zeros of the Higgs field
  on the vertices of a polyhedron with icosahedral symmetry.
  A more detailed picture of the Higgs field is provided in
  Figure~\ref{fig-antibag}, where we plot the length of the
  Higgs field $|\Phi|$ as a function of geodesic distance from the
  origin $\rho$ along a radial half-line that passes through a
  vertex of the polyhedron (black curve) and a face centre of
  the polyhedron (yellow curve). The blue curve is the
  spherical average of
  $|\Phi|$, obtained by integrating over the angular coordinates.

It is immediately clear from Figure~\ref{fig-antibag} that an
abelian magnetic bag does not provide a good description of this
large charge hyperbolic monopole, because the
length of the Higgs field does not remain close to zero in a region
that could be associated with the interior of an abelian bag.
Furthermore, suggested generalizations \cite{LW,Man}, in which
the length of the Higgs 
field is assumed to be a non-zero constant in the interior
of the bag, are also not appropriate
here, as $|\Phi|$ has a significant $\rho$ dependence.

Figure~\ref{fig-antibag} reveals that
the best that any spherical bag description
could hope to achieve is an approximation to the spherical average of
$|\Phi|$. This is because there is a substantial angular variation
of $|\Phi|$ on the sphere that contains most of the zeros of the
Higgs field.
As $v=\frac{1}{2}$, the spherical abelian magnetic bag 
(\ref{outside}) is given by
  \be
  \phi  =\frac{1}{2}(N+1-N\coth\rho).
  \label{exterior}
  \ee
  As we shall see, this does provide a good description of the
  spherical average of $|\Phi|$ in the exterior of a suitable bag, but
  clearly it fails in the interior.

  A key observation from Figure~\ref{fig-antibag} is that the monopole
  appears to be spherically symmetric in a large region around the
  origin, that we identify as the interior of our new bag.
  Although a spherical abelian description is not valid
  in the interior, it turns out
  that a spherical non-abelian solution of the Bogomolny equation is
  an excellent approximation in this region.

  Let $\theta,\chi$ be the usual angular coordinates on the sphere, as in
  (\ref{spherical}). The standard spherical hedgehog ansatz,
  in radial gauge $A_\rho=0$, is given by
  \bea
  \Phi&=&ih(\sin\theta(\tau_1\cos\chi+\tau_2\sin\chi)+\tau_3\cos\theta),\\
  A_\theta&=&\frac{i}{2}(k-1)(\tau_1\sin\chi-\tau_2\cos\chi),\\
  A_\chi&=&\frac{i}{2}(k-1)\sin\theta((\tau_1\cos\chi+\tau_2\sin\chi)\cos\theta-\tau_3\sin\theta),
  \eea
  where $\tau_i$ are the Pauli matrices and $h,k$ are radial profile
  functions that depend only on $\rho$.
  Substituting this hedgehog ansatz into the Bogomolny equation (\ref{bog})
  yields the following ordinary differential equations for $h(\rho)$ and
  $k(\rho)$
  \be
  \frac{dh}{d\rho}=\frac{1-k^2}{2\,{\rm sinh}^2\rho}, \qquad \qquad \qquad
  \frac{dk}{d\rho}=-2hk.
  \label{radbog}
  \ee
  Regularity at the origin imposes the boundary conditions $h(0)=0$ and
  $k(0)=1$. Requiring the correct asymptotic value for the length of
  the Higgs field imposes the condition
  \be
  |\Phi|=|h|\to v=\frac{1}{2} \quad\mbox{ as } \quad \rho\to\infty.
  \label{bc}
  \ee
The standard 1-monopole solution of (\ref{radbog})  
is given by
  \be
  h={\rm coth}(2\rho)-\frac{1}{2}{\rm coth}\rho,
  \qquad\qquad\qquad
  k={\rm sech}\rho.
  \label{1mon}
  \ee
  This solution has the small $\rho$ expansion
  $h=\frac{\rho}{2}+{\cal O}(\rho^3)$, and the
  fact that the coefficient of the linear term is positive corresponds
  to a zero of the Higgs field at the origin with multiplicity $+1$.
Note that $k\to 0$ as $\rho\to\infty,$ which is a finite energy requirement.

There is another solution of (\ref{radbog})
that satisfies the regularity conditions at the origin and
the boundary condition (\ref{bc}). It is given by
  \be
  h=\frac{1}{2\rho}-\frac{1}{2}{\rm coth}\rho,
  \qquad\qquad\qquad
  k=\frac{{\rm sinh}\rho}{\rho}.
  \label{inside}
  \ee
  This solution does not have a finite charge $N$ because
  $k\not\to 0$ as $\rho\to\infty,$
   but rather it grows without bound. However, it is a perfectly regular solution for any finite value of
  $\rho.$
  The small $\rho$ expansion of this solution gives
  $h=-\frac{\rho}{6}+{\cal O}(\rho^3)$
  and hence there is an anti-zero of the Higgs field at the origin,
  because the coefficient of the linear term is negative.

  The scalar field $\phi$, that approximates the
  spherical average of $|\Phi|$,
  is obtained for our new magnetic bag
  by taking the non-abelian solution (\ref{inside})
  in the interior of the bag and the abelian solution
  (\ref{exterior}) in the exterior of the bag.
  Explicitly,
  \be
  \phi=\begin{cases}
    \frac{1}{2}{\rm coth}\rho-\frac{1}{2\rho}
  & \text{for } \  0\le\rho\le\rho_\star\\
  \frac{1}{2}(N+1-N{\rm coth}\rho)
  & \text{for } \  \rho>\rho_\star,
  \end{cases}
  \label{newbag}
  \ee
  where 
  the bag radius $\rho_\star$ is determined in terms of the magnetic charge
  $N$ by requiring that $\phi$ is continuous
  at $\rho=\rho_\star.$ The result is 
  \be
  N=  \frac{e^{2\rho_\star}-2\rho_\star-1}{2\rho_\star}.
  \ee
  For $N=371$ this gives $\rho_\star\approx 4$ and the
  associated new magnetic
  bag (\ref{newbag}) is shown as the
  red curve in Figure~\ref{fig-antibag}.
  This plot demonstrates that the new magnetic bag provides an
  excellent approximation to the spherical average of this
  hyperbolic monopole.

  We have performed a similar comparison
  for a range of large charge strawberry flavour
  hyperbolic monopoles obtained from solutions of the Thomson problem,
  with the result that the same level of excellent agreement is found.
  Not only does this demonstrate the success of our new magnetic bag
  approximation, but it elucidates the nature of monopole anti-zeros.
  Until now, this has been somewhat of a mysterious issue, but now we
  see that a monopole with an anti-zero is simply making use of a
  previously overlooked spherically symmetric solution of the
  Bogomolny equation. There is a similar spherically symmetric solution
  of the Bogomolny equation in $\mathbb{R}^3$,
  satisfying the regularity conditions at the origin but not the
  finite energy condition at infinity, so this new understanding of
  monopole anti-zeros extends to the Euclidean setting too.

  The observant reader may wonder why we chose to impose the
  $\rho\to\infty$ boundary
  condition \eqref{bc} on the solution used for the
  interior of the bag, given that the bag approximation \eqref{newbag}
  only utilises this solution in the finite range $[0,\rho_\ast]$.  Our
  justification is that the solution \eqref{inside} fits
  the exact monopole fields.  We note however that the system
  \eqref{radbog} has many solutions with an anti-zero at $\rho=0$ other
  than \eqref{inside}; the fact that the particular solution
  \eqref{inside} fits all available strawberry flavour
  monopoles may be a consequence of
  working within the JNR ansatz.

  \section{Conclusion}\quad\
  The abelian magnetic bag, describing a large number of
  coincident non-abelian BPS monopoles, has been extended
  to hyperbolic space and its properties investigated
  in detail. In particular, we have
  made comparisons with exact solutions of the Bogomolny equation
  containing hundreds of monopoles.
  This is the main reason for moving to the hyperbolic setting, as
  such exact solutions are not available for comparison in
  Euclidean space. Our results show a good agreement for charge $N$
  monopoles with a single zero of the Higgs field (of multiplicity $N$)
  and we have derived a Nahm transform for the associated abelian
  magnetic bag from the large $N$ limit of the discrete Nahm equation
  for hyperbolic monopoles.
  However, for monopoles with more than $N$ zeros of the Higgs field we find
  that the abelian magnetic bag is not a good description, but must be
  supplemented by a non-abelian interior for the bag, which we are
  able to describe in detail. This provides a new understanding of the
  structure of monopole anti-zeros.

\section*{Acknowledgements}
This work is funded by the EPSRC grant EP/K003453/1 and 
the STFC grant ST/J000426/1.

\end{document}